\documentclass[preprint]{revtex4}
\usepackage{graphicx}
\newcommand{\be}{\begin{equation}}
\newcommand{\bea}{\begin{eqnarray}}
\newcommand{\ee}{\end{equation}}
\newcommand{\eea}{\end{eqnarray}}
\newcommand{\eps}{\epsilon}
\newcommand{\om}{\omega}

\begin{document}
\title{ Resonant phenomena in slowly perturbed elliptic billiards
} \author{A.P. Itin, A.I. Neishtadt}
 \email{aneishta@iki.rssi.ru} \affiliation{Space Research
Institute, Russian Academy of Sciences, Profsoyuznaya Str. 84/32,
117997 Moscow, Russia. }

\begin{abstract}
We consider an elliptic billiard whose shape slowly changes.
During slow evolution of the billiard certain resonance conditions
can be fulfilled. We study the phenomena of capture into a
resonance and scattering on a resonance which lead to the
destruction of the adiabatic invariance in the system.
\end{abstract}

\maketitle

Billiard systems are important models in different fields of
physics and the theory of dynamical systems
\cite{Ar,ZS,KoTr,Exp2,Nir}. Recently the billiards with varying
parameters became the object of interest \cite{kol}. In the
present paper we consider effects of scattering on resonances  and
capture into a resonance in the dynamics of a particle in an
elliptic billiard with slowly varying parameters. We use a scheme
of analysis of resonant phenomena in Hamiltonian systems
possessing slow and fast variables \cite{kluwer}; this scheme is a
Hamiltonian version of a general scheme \cite{general1,general2}.
These methods were designed and corresponding theorems were proved
for smooth systems. In the recent paper on slowly perturbed
rectangular billiards \cite{Rect} it was shown that the methods
can be also applied adequately  for billiard systems which possess
discontinuities. The present paper gives another example of a
billiard system possessing effects of capture into a resonance and
scattering on a resonance.

Consider a particle moving in an elliptic billiard. Let $r_1,r_2$
be the particle distances from foci (points $O_1, O_2$
correspondingly), and 2c is the distance between the foci. The
Hamiltonian of the system in elliptic coordinates $\xi = r_1+r_2,
\quad \eta=r_1-r_2$ has the following form \cite{Ar} : \be H_0= 2
p_{\xi}^2 \frac{\xi^2-4c^2}{\xi^2-\eta^2} + 2 p_{\eta}^2 \frac{4
c^2- \eta^2}{\xi^2-\eta^2} + U(\xi, a), \ee where $U(\xi,a)$ is
the potential of the billiard wall: \bea U(\xi,a)= \left\{
\begin{array}{rcl} \infty, \quad \mbox{if  } \xi> 2a
\\0 , \quad \mbox{if  }  \xi< 2a
 \end{array}\right.
\eea We investigate the problem of an elliptic billiard with
slowly changing parameters $c=c(\eps t), \quad a=a(\eps t)$ which
slowly rotates with angular velocity $\omega$. Let us start with
cartesian coordinates ($x,y$) such that the ellipse foci are
located in $(c,0)$ and $(-c,0)$. Let ($p_x,p_y$) be momenta
conjugated to ($x,y$). Changing to new variables ($P_u,u,P_v,v$)
by means of a canonical transformation with a generating function
$W=c(p_x \cosh v \cos u - p_y \sinh v \sin u)$ (see Ref.
\cite{Thiele-Burrau})
we get the Hamiltonian \bea H= \frac{P_v^2+P_u^2}{c^2(\cosh 2
v-\cos 2u)}+  \frac{\eps c'}{c}  \cdot \frac{P_u \sin 2u - P_v
\sinh 2v}{\cosh 2v-\cos 2u}+ \omega \cdot \frac{P_u \sinh 2v + P_v
\sin 2u}{\cosh 2v-\cos 2u}+ {\cal U}(v,a/c),\nonumber\\
\label{hinitial} \eea where \bea {\cal U}(v,a/c)= \left\{
\begin{array}{rcl} \infty, \quad \mbox{if } \cosh v> a/c
\\0 , \quad \mbox{if  }  \cosh v< a/c
 \end{array}\right.
\eea

 New coordinates relate to the elliptic coordinates by means
 of the following formulas: \be \xi=2c \cosh v, \quad \eta= 2 c\cos u.
 \ee
 Let us introduce the variable
$P_t$ which is canonically conjugated to time $t$, and the
variable $\tau= \eps t$. Now we can consider the system as the
autonomous three-dimensional
system with Hamiltonian: $H'(P_u,u,P_v,v, P_t,\tau)=H+P_t$.
Canonically conjugated pairs are $(P_u,u)$, $(P_v,v)$, and $(P_t,
\eps^{-1}\tau)$.
This Hamiltonian system is conservative. Let us consider its
dynamics on some level of the Hamiltonian $H'=E$. Denote
$k=2(E-P_t)$ (in the unperturbed system $k>0$, since the
Hamiltonian (\ref{hinitial}) is greater than $0$).
 By means of simple canonical transformation we can change to Hamiltonian $
H=H'-E$ and then introduce new time variable $t'$ such that
$dt/dt'$=$c^2(\cosh 2 v-\cos 2 u)$ \cite{Thiele-Burrau}.  Since
the value of the Hamiltonian $H=H'-E$ is equal to $0$, after
introducing new time variable $t'$ new Hamiltonian ${\cal H}$ is
obtained by multiplying the Hamiltonian $H$ by $c^2(\cosh 2 v-\cos
2 u)$.

So we get the following Hamiltonian: \bea {\cal H} &=& P_v^2 - c^2
k \cosh^2 v + P_u^2 + c^2 k \cos^2 u
+ {\cal U}(v,a/c) + \eps c' c \left[ P_u \sin 2u -  P_v \sinh 2 v \right] +\nonumber\\
&+&\omega c^2 \left[ P_u \sinh 2v +  P_v \sin 2 u \right] =
 {\cal F}_0 + \eps {\cal F}_1 \equiv 0. \label{fin}
\eea

The Hamiltonian (\ref{fin}) describes the dynamics of a particle
in the elliptic billiard perturbed by slow deformation and
rotation. The unperturbed problem ($\tau$=const,  $\eps =\omega=
0$) is integrable. In addition to conserved energy there is
another constant of motion $c_1=c_u=c_v$: \bea
P_v^2 - c^2 k \cosh^2 v =c_v, \label{oscuv}\\
P_u^2 + c^2 k \cos^2 u =-c_u. \nonumber\eea
We introduced variables $c_u,c_v$ which are equal to each other in
the unperturbed system and slightly differ in the perturbed
system. The unperturbed system can be regarded as two uncoupled
oscillators whose phase portraits are presented in Fig. 1.
One can introduce action-angle variables $(I_u,\phi_u,I_v,\phi_v)
$ by means of canonical transformation with a generating function
$S = S(I_u,I_v,u,v,P_t,\tau)$ which contains $\tau,P_t$ as
parameters. In the new variables the unperturbed Hamiltonian
${\cal F}_0$ transforms to $
 {\cal H}_0 = {\cal H}_0(I_u,I_v,P_t,\tau)= {\cal
H}_u(I_u,P_t,\tau)+{\cal H}_v(I_v,P_t,\tau). $
 The function $S$ has the form \be S=
\int_{v_0}^{v}\sqrt{c_v+c^2k \cosh^2 x} dx +
\int_{u_0}^{u}\sqrt{-c_u-c^2k \cos^2 y} dy. \label{s} \ee
Actions of the system have the following form (see also
\cite{Richter1,Richter2}): \bea
\mbox{Case 1} &:& c^2k+c_u>0. \nonumber\\
I_v =\frac{1}{\pi}  \int_0^{\cosh^{-1}(a/c)} dv \sqrt{c_v+c^2k
\cosh^ 2 v} &=& \nonumber\\ = \frac{1}{\pi}
\Biggl[\frac{c_v+kc^2}{\sqrt{kc^2}} F(\alpha,m)- \sqrt{ k c^2} E
(\alpha,m) &+& ka \sqrt{\frac{ a^2-c^2}{c_v+ka^2}} \Biggr],
\nonumber\\
I_u = \frac{1}{\pi}
\int_{\pi/2}^{\cos^{-1}{-\sqrt{\frac{-c_u}{c^2k}}}} du
\sqrt{-c_u-c^2k \cos^2 u} &=& \frac{1}{\pi} \Biggl[-
\frac{c_u+kc^2}{\sqrt{kc^2}} {\bf K}(m) + \sqrt{kc^2}{\bf E}(m)  \Biggr] , \\
\mbox{where } m &=& -\frac{c_u}{ c^2 k}, \quad \alpha =
\arcsin \sqrt{\frac{k(a^2-c^2) }{ka^2+c_u}}.  \nonumber\\
\nonumber\\
\mbox{Case 2} &:& c^2 k+c_u<0. \nonumber\\
I_v= \frac{1}{\pi}
\int_{\cosh^{-1}\sqrt{\frac{-c_v}{c^2k}}}^{\cosh^{-1}(a/c)} dv
\sqrt{c_v+c^2k \cosh^2 v} &=& \frac{1}{\pi} \left[ a
\sqrt{\frac{a^2 k
+c_v}{a^2-c^2}} -\sqrt{-c_v} E \left(\beta, \frac{1}{m} \right) \right],  \nonumber\\
I_u = \frac{1}{\pi}  \int_{\pi/2}^{\pi} du \sqrt{-c_u-c^2k \cos^2
u}
&=& \frac{\sqrt{-c_u}}{\pi} {\bf E}(\frac{1}{m}), \qquad \\
 \mbox{where }  \beta=\arcsin \sqrt{\frac{
 ka^2+c_u}{k(a^2-c^2)}}.
  \nonumber
\eea

Frequencies of the system have the form \bea
\mbox{case 1}&:& \nonumber\\
\omega_v= \frac{\partial {\cal H}_0}{\partial I_v} =
\frac{1}{\frac{\partial I_v}{\partial c_v}} = \frac{2 \pi
\sqrt{c^2k}}{  F( \alpha,m) }, \qquad \omega_u= \frac{\partial
{\cal H}_0}{\partial I_u} = - \frac{1}{\frac{\partial
I_u}{\partial c_u}} &=&  \frac{2 \pi \sqrt{c^2k} }{{\bf K}(m )}.
\nonumber\\
\nonumber\\
\mbox{case 2}&:& \nonumber\\
\omega_v = \frac{2 \pi \sqrt{-c_v}}{F (\beta, \frac{1}{m}) },
\qquad \omega_u &=& \frac{2 \pi \sqrt{-c_u}}{ {\bf K}
\left(\frac{1}{m} \right)}. \nonumber \eea

Now consider the perturbed problem $(\eps \ne 0)$, following Ref.
\cite{kluwer}. Let us make in the system with Hamiltonian
(\ref{fin}) the canonical transformation of the variables \be
(P_u,u,P_v,v,P_t,\tau) \to (\bar{I}_u,\bar{\phi}_u,\bar{I}_v,\bar
{\phi}_v, \bar P_t, \bar \tau) \ee determined by the generating
function \be S_2 = \bar P_t \eps ^{-1} \tau + S( \bar{I}_u, {\bar
I}_v,u,v,{\bar P}_t,\tau). \ee

Formulas for the transformation of the variables have the form
\bea \bar \phi_{\alpha} = \frac{\partial S}{\partial \bar
I_{\alpha}},
\quad \alpha=u,v, \nonumber\\
P_v= \frac{\partial S}{\partial v}, \quad
P_u= \frac{\partial S}{\partial u},  \\
P_t =\bar P_t + \eps \frac{\partial S}{\partial \tau}, \quad \bar
\tau = \tau + \eps \frac{\partial S}{\partial \bar P_t}.
\nonumber\eea Hamiltonian (\ref{fin}) in the new variables is as
follows: \bea {\cal H}= {\cal H}_0(\bar I_u,\bar I_v,\bar P_t,\bar
\tau) &+& \eps {\cal H}_1(\bar I_u,\bar I_v,\bar \phi_u,\bar
\phi_v, \bar P_t,\bar \tau)
+ O(\eps^2), \nonumber\\
{\cal H}_1 &=& {\cal F}_1 + \frac{\partial {\cal F}_0}{\partial
P_t} \frac{\partial S}{\partial \tau}-\frac{\partial {\cal
H}_0}{\partial \tau} \frac{\partial S}{\partial \bar P_t} .\eea
The variables $(\bar I_u,\bar \phi_u,\bar I_v,\bar \phi_v, \bar
P_t, \bar \tau)$ are $O(\eps)$- close to the variables
$(I_u,\phi_u,I_v,\phi_v, P_t, \tau)$. Henceforth the bar over the
new variables is omitted and the new Hamiltonian is \be {\cal H}=
{\cal H}_0(I_u,I_v,P_t,\tau)+ \eps {\cal
H}_1(I_u,I_v,\phi_u,\phi_v,P_t,\tau) + O(\eps^2). \label{last} \ee
The differential equations of the motion have the form \bea \dot
I_{\alpha} = - \eps \frac{\partial {\cal H}_1}{\partial
\phi_{\alpha}} +O(\eps^2), \quad \dot \phi_{\alpha} =
\omega_{\alpha}(I_u,I_v,P_t,\tau) + \eps \frac{\partial {\cal
H}_1}{\partial I_{\alpha}} +O(\eps^2),
 \qquad \alpha=u,v,  \label{ur}\\
\dot P_t= -\eps \frac{\partial {\cal H}_0}{\partial \tau}- \eps^2
 \frac{\partial {\cal H}_1}{\partial \tau} + O(\eps^3), \quad
 \dot \tau = \eps \frac{\partial {\cal H}_0}{\partial P_t}+ \eps^2
 \frac{\partial {\cal H}_1}{\partial P_t} + O(\eps^3).\nonumber
\eea Averaging of the right hand sides of (\ref{ur}) over
$\phi_{\alpha}$ and discarding terms $O(\eps^2)$ gives an averaged
system \be \dot I_{\alpha}=0, \quad \dot P_t= -\eps \frac{\partial
{\cal H}_0}{\partial \tau}, \quad \dot \tau = \eps \frac{\partial
{\cal H}_0}{\partial P_t}.   \label{app} \ee Approximation
(\ref{app}) is called an adiabatic approximation \cite{general2}.
Trajectories of the system (\ref{app}) are called adiabatic
trajectories. In the adiabatic approximation $I_{u,v}=
\mbox{const}$. The adiabatic approximation breaks down in a
vicinity of a resonant surface in the phase space where the
resonance condition $k_u \omega_u+k_v \omega_v =0 $ is fulfilled
($k_u, k_v$ are integers, $k_u^2+k_v^2 \ne 0 $). Although resonant
surfaces are dense in the phase space of our three-dimensional
Hamiltonian system, for small $\eps$ only finite number of
low-order resonances are important (the order of a resonance is $k
= |k_u|+|k_v|)$ \cite{kluwer}.

 In the exact (non-averaged) system variables $I_{\alpha}$ are
approximate adiabatic invariants, i.e. they are well-conserved in
a large area of phase space (far from resonant surfaces). Near a
resonant surface of a given resonance the system (\ref{last}) can
be transformed into standard "perturbed pendulum-like system"
form. Details can be found in Ref. \cite{kluwer}.

During the motion near ($k_u,k_v$) - resonance the magnitude of
$J=-k_v I_u+ k_u I_v$ is approximately conserved.
Far from resonances the magnitudes of $I_u, I_v$ are approximately conserved.

So, a phase point in the $(I_u,I_v,P_t,\tau)$ - space moves in a
following way: while it is far from low-order resonance surfaces
$k_u \om_u(I_u,P_t,\tau)+ k_v \om_v(I_v,P_t, \tau)=0$, it moves in
a vicinity of an adiabatic curve $I_{u,v}=$const. When it
approaches a resonant surface (enters a resonant zone), it leaves
the adiabatic curve and can be either captured into the resonance
or scattered on the resonance. In the former case it continues its
motion in the vicinity of the resonant surface until become
ejected from the resonance, whereas in the latter case its
adiabatic invariants $I_u,I_v$ undergo jumps $\sim \sqrt{\eps}$
and after crossing the resonant zone the particle continues its
motion along another adiabatic curve. Similar dynamics was
investigated recently in Refs.
\cite{Rect,surfatron,esurfatron,Itbc}. Corresponding dynamics of
the particle in the coordinate space $(x,y)$ can be described as
follows. In the adiabatic approximation each segment of the
particle trajectory in a coordinate system rotating with the
billiard tangents a confocal quadrics (caustics) determined by
conditions $I_{u,v}$ = const. The caustics slowly evolves in
accordance with evolution of the billiard parameters. In the exact
system, while a phase point is far from low-order resonance
surfaces, each segment of the particle trajectory tangents a
confocal quadrics which is close to the caustics determined by
$I_{u,v}$ = const. While captured in a $(m,n)$ resonance, each
segment of the particle trajectory tangents a confocal quadrics
which is close to the quadrics determined by conditions $J$ =
$-nI_u+mI_v$ = const, $\omega_u/\omega_v =-n/m$ = const.

Consider the case of the billiard whose parameters $a,c$ evolve
periodically in time. If an adiabatic trajectory crosses a
resonant surface, it crosses the surface at the same point
periodically in time. A phase point which moves near this
adiabatic trajectory also crosses the resonant surface repeatedly.
Accumulation of changes of $I_u,I_v$ due to multiple passages
through resonances leads to the destruction of the adiabatic
invariance in the system \cite{kluwer}.

It should be noted that the billiard could be deformed in such a
way that there will be no passages through resonances and the
dynamics will be determined by KAM theory  \cite{Ar}. Let the
parameters $a$ and $c$ be  changed in such a way that $a/c=$const.
In the system (\ref{fin}) one can change from variables
$(P_t,\tau)$ to $( {\tilde P}_t =c^2k, {\tilde  \tau}={\tilde
\tau}( \tau)=\int_{\tau_0}^{\tau} \frac{d\tau}{c^2}$) using simple
canonical transformation which does not change other variables.
Then, Hamiltonian ${\cal H}_0$ depends only on actions
$I_u,I_v,{\tilde P}_t$ and KAM theory could be applied. This means
that most of the phase space ($I_u,I_v,{\tilde P}_t$) is filled up
by invariant tori which are close to the tori $I_u,I_v,{\tilde
P}_t$=const.

 We performed numerical investigations of the system. The
results are shown in figures 2-6. In figures 2-3 the results of
numerical investigations of slowly deforming elliptic billiards
are shown. Semi-axis of a billiard, $d_a$ and $d_b$, were being
changed periodically with time. Jumps of the adiabatic invariant
$I_v$ on a (2,1) resonance are shown in Fig. 2. A single jump of
the adiabatic invariant is shown in Fig. 3. In Figures 4-6 the
results of numerical investigations of slowly rotating and
deforming billiards are presented. In Fig. 4 the capture of a
phase point into a (2,1) resonance is shown. The captured point
moves along a resonant curve until it escapes from the resonance.
In Fig. 5 a phase point that were initially captured into
resonance remains captured forever. Dynamics of adiabatic
invariant $J$ of that system is shown in Fig. 6. In Fig. 7 we
depicted results of numerical investigation of a deforming
billiard with $\omega=0$, $a/c=$ const. Coordinates (x,y) of a
particle at the moments of collisions with a wall are shown. It is
easy to see that dynamics is regular (compare that picture with
Fig. 8, where $a/c \ne$ const).
\\

This work was supported in part by grants RFBR 00-01-00538, RFBR
00-15-96146, and INTAS  00-221. A.I. Neishtadt acknowledges
support from the program "Integration"-B0053.





\begin{figure*}
{\includegraphics[width=8cm]{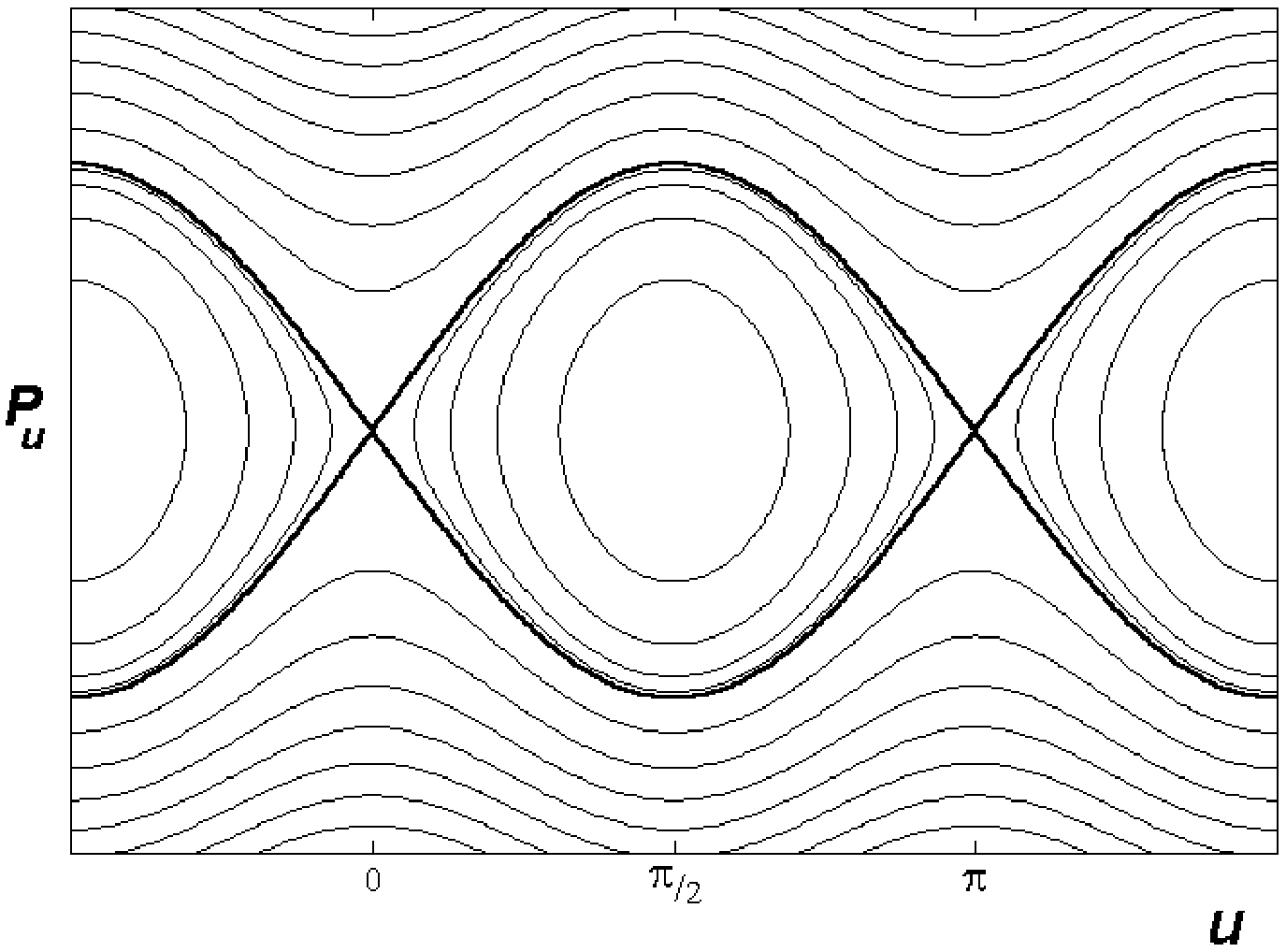}(a)}
 {\includegraphics[width=8cm]{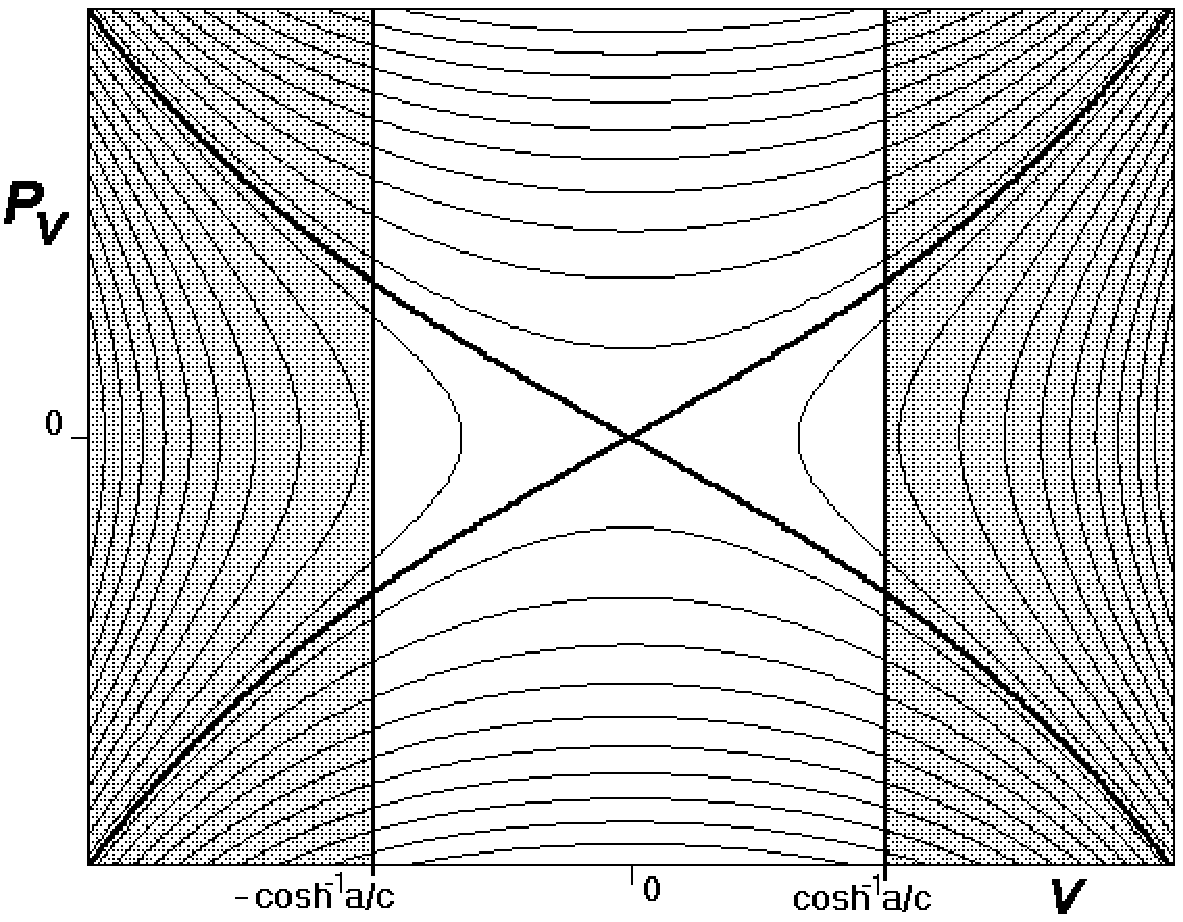}(b)}
 \caption{
 Phase portraits of the oscillators (7).
a) $(P_u,u)$ plane. If $c^2 k+c_u>0 $ (case 1), the motion in the
$(P_u,u)$ plane is libration. If $c^2 k+c_u<0 $ (case 2), the motion is
rotation.
 b) $(P_v,v)$ plane. Vertical lines $v= \pm
\cosh^{-1} a/c$ correspond to a potential wall of the billiard. The
region between the lines corresponds to phase points moving in the
billiard. Shaded region outside the lines corresponds to phase points
outside the billiard (we do not consider their motion).
 In the case 1 ($c^2 k +c_v>0$)
a phase point which begin to move from a point on the "left" wall with
$P_v>0$ goes to the right until a collision with the "right" wall, then
jumps down instantaneously to the point located symmetrically about the
line {$P_v=0$}, then moves to the left until collision with the "left"
wall, then jumps to the initial point, and so on. In the case 2 ($c^2 k
+c_v<0$) a phase point which begin to move from a point on the "left"
wall with $P_v>0$ goes downwards until a collision with the same wall,
then jumps to the point located symmetrically about the line {$v=0$},
then moves upwards until a collision with the "right" wall, then jumps
to the initial point, and so on. }
\end{figure*}



\begin{figure*}
\includegraphics[width=8cm]{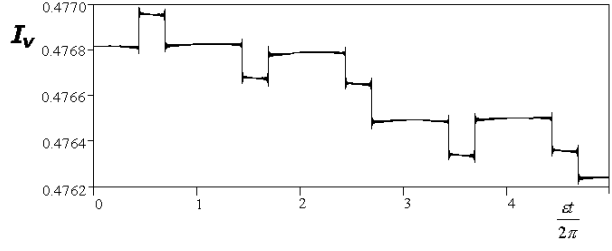}
 \caption{
 Jumps of the adiabatic invariant $I_v$ on
(2,1) resonance. Parameters of the system:
 $\eps=8*10^{-5}$, $\omega=0$, $d_i$
change harmonically with time: $d_a=d_1 (1+A_1 \cos(\eps t))$,
$d_b=d_2(1+A_2 \cos(\eps t+\phi))$, where $A_1=0.3$, $A_2=0.15$,
$d_1=2$, $d_2=1$, $\phi=\pi/3$. }
\end{figure*}

\begin{figure*}
\includegraphics[width=8cm]{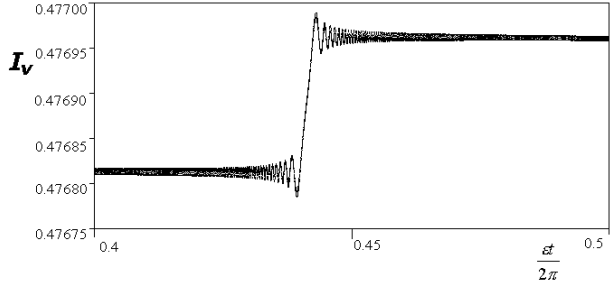}
 \caption{
 A single jump of the adiabatic invariant on a
(2,1) resonance. Parameters are the same as in fig 2. }
\end{figure*}

\begin{figure*}
\includegraphics[width=8cm]{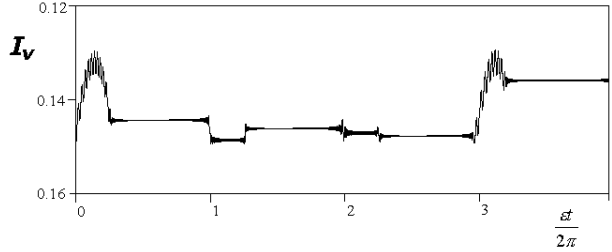}
 \caption{
 Jumps of adiabatic invariant $I_v$ and
captures of a phase point into a (2,1) resonance. The captured
point moves along a resonant curve until it escapes from the
resonance. Parameters of the system:
 $\eps=10^{-3}$, $\omega=2*10^{-4}$, $d_i$ change harmonically with time, as in figures 2-3 except that
$A_2=0.2$ }
\end{figure*}

\begin{figure*}
{\includegraphics[width=8cm]{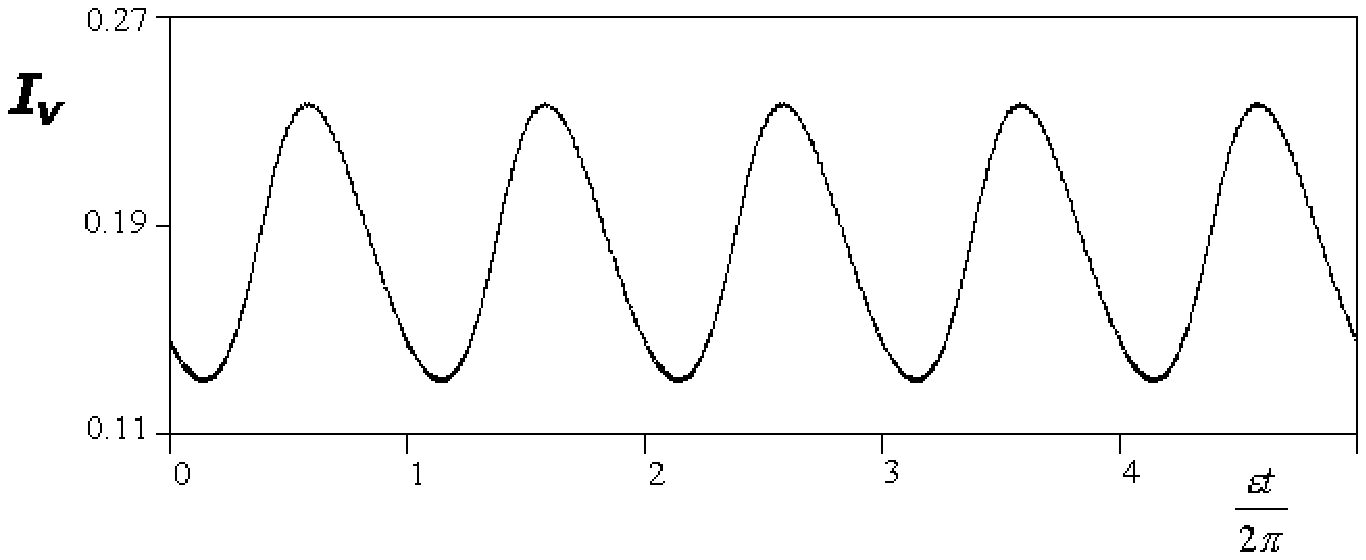}(a)}
{\includegraphics[width=8cm]{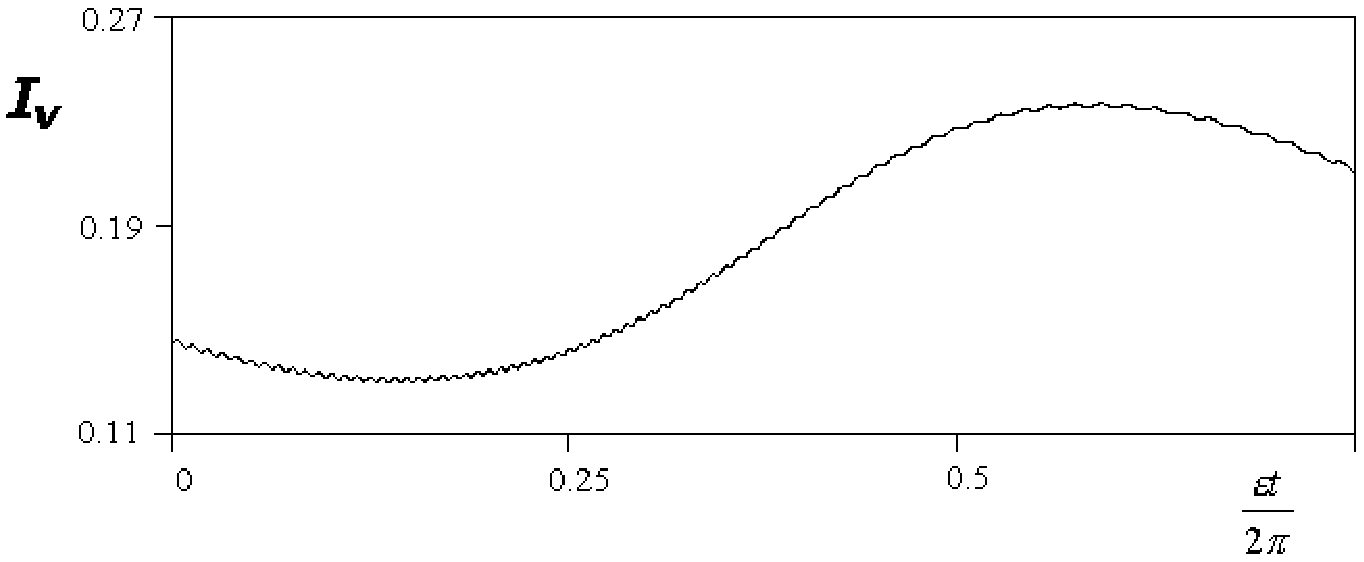}(b)}
 \caption{Dynamics of the adiabatic invariant $I_v$ of a
phase point captured into the resonance (2,1). The phase point remains
captured forever. Parameters are the same as in Fig. 4 (initial
conditions are different). In Fig. a) and b) scales are different. }
\end{figure*}

\begin{figure*}
{\includegraphics[width=8cm]{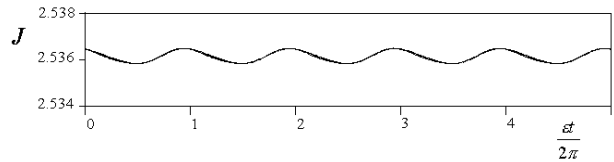}}
 \caption{
Dynamics of the adiabatic invariant $J$. Parameters and initial
conditions are the same as in Fig 5.}
\end{figure*}

\begin{figure*}
{\includegraphics[width=8cm]{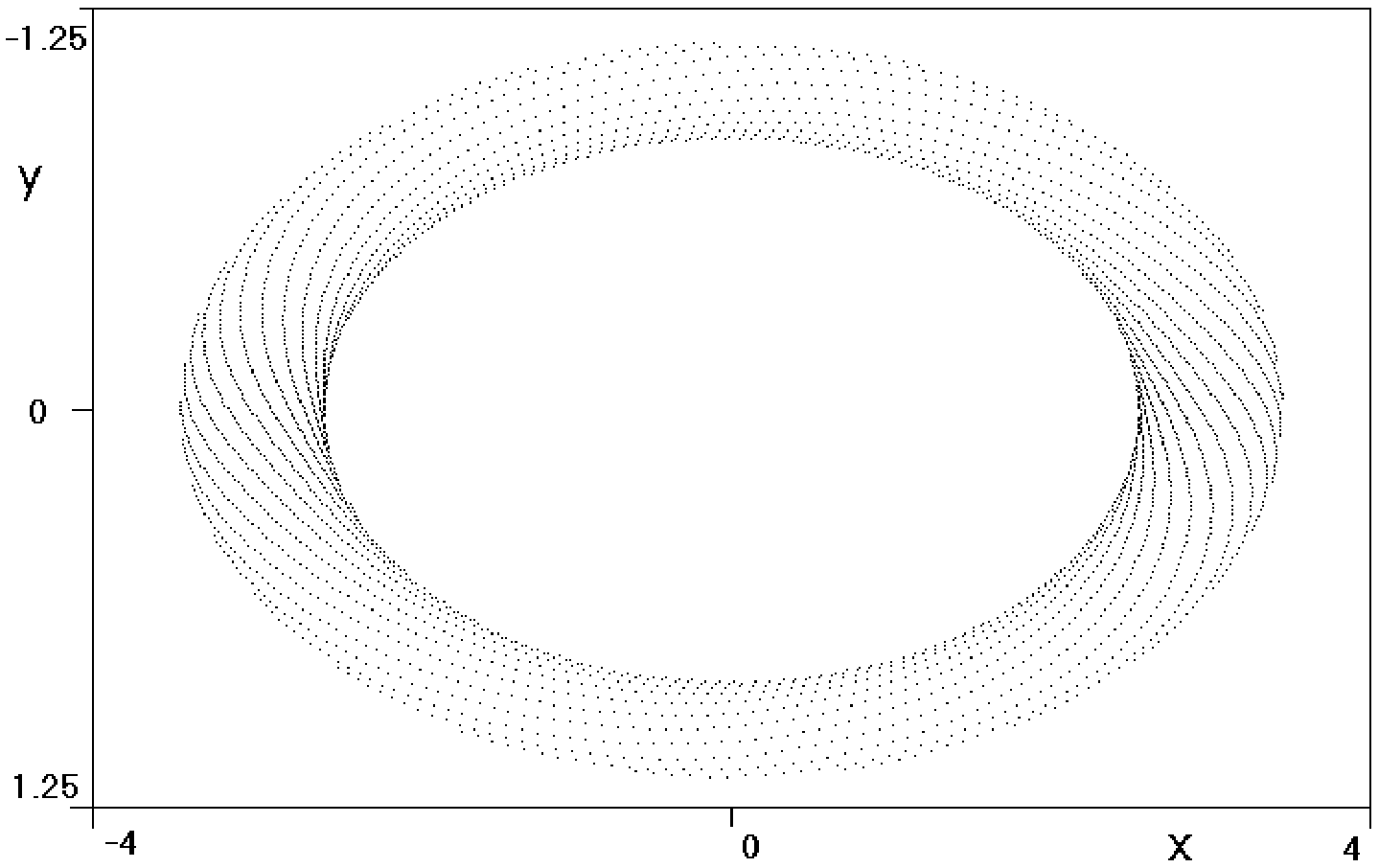}(a)}
{\includegraphics[width=8cm]{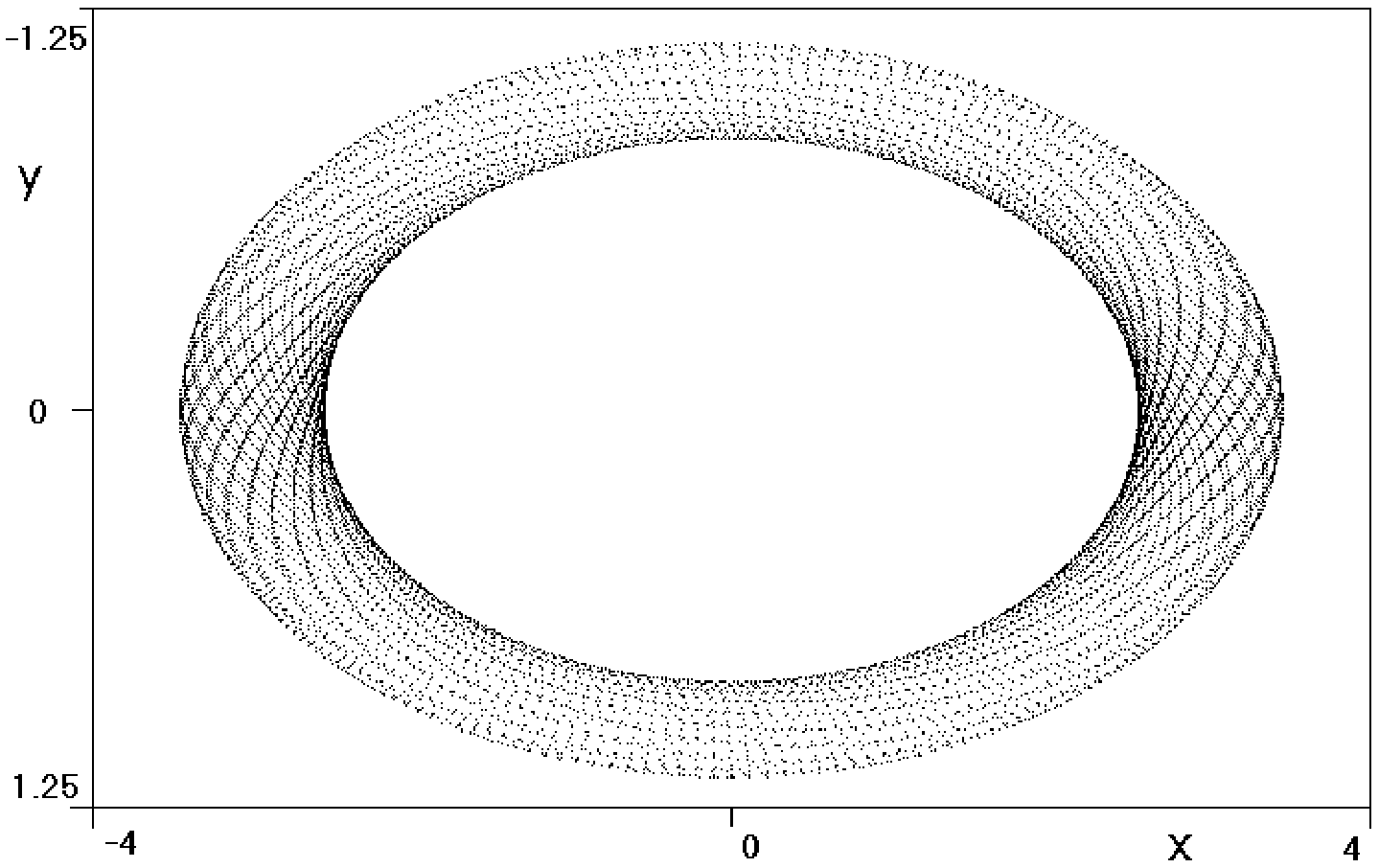}(b)}
 \caption{Regular dynamics of the system with
$a/c=$const. Coordinates $(x,y)$ of a particle at the moments of
collisions with a wall are shown. Parameters of the system:
$\eps=0.001,$ $\omega=0$, $A_1=0.15$, $A_2=0.15$, $d_1=3$, $d_2=1$,
$\phi=0$. a) Time of the integration is one-half of the slow period.
b)Time of the integration is equal to 2 slow periods.}
\end{figure*}

\begin{figure*}
{\includegraphics[width=8cm]{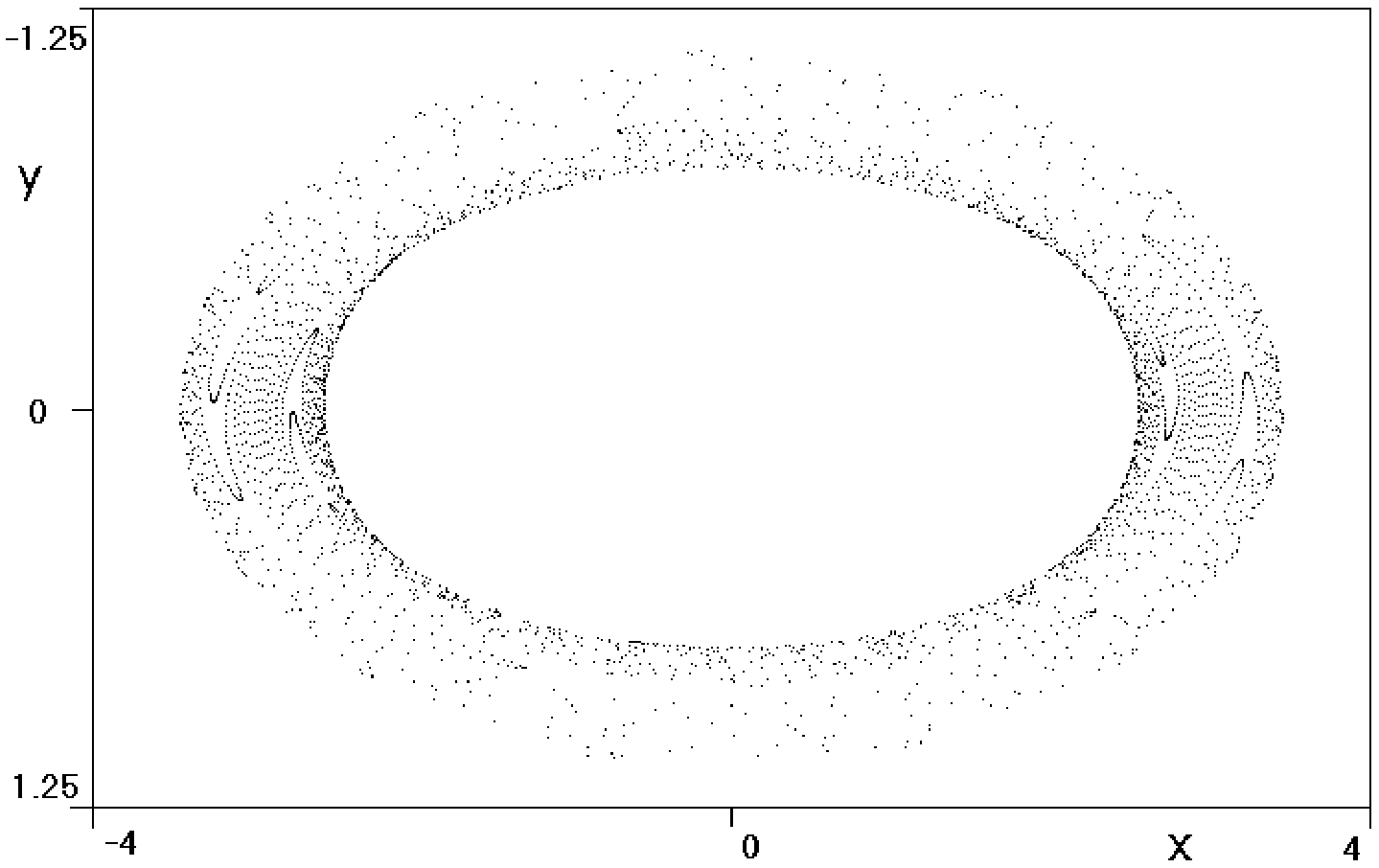}(a)}
{\includegraphics[width=8cm]{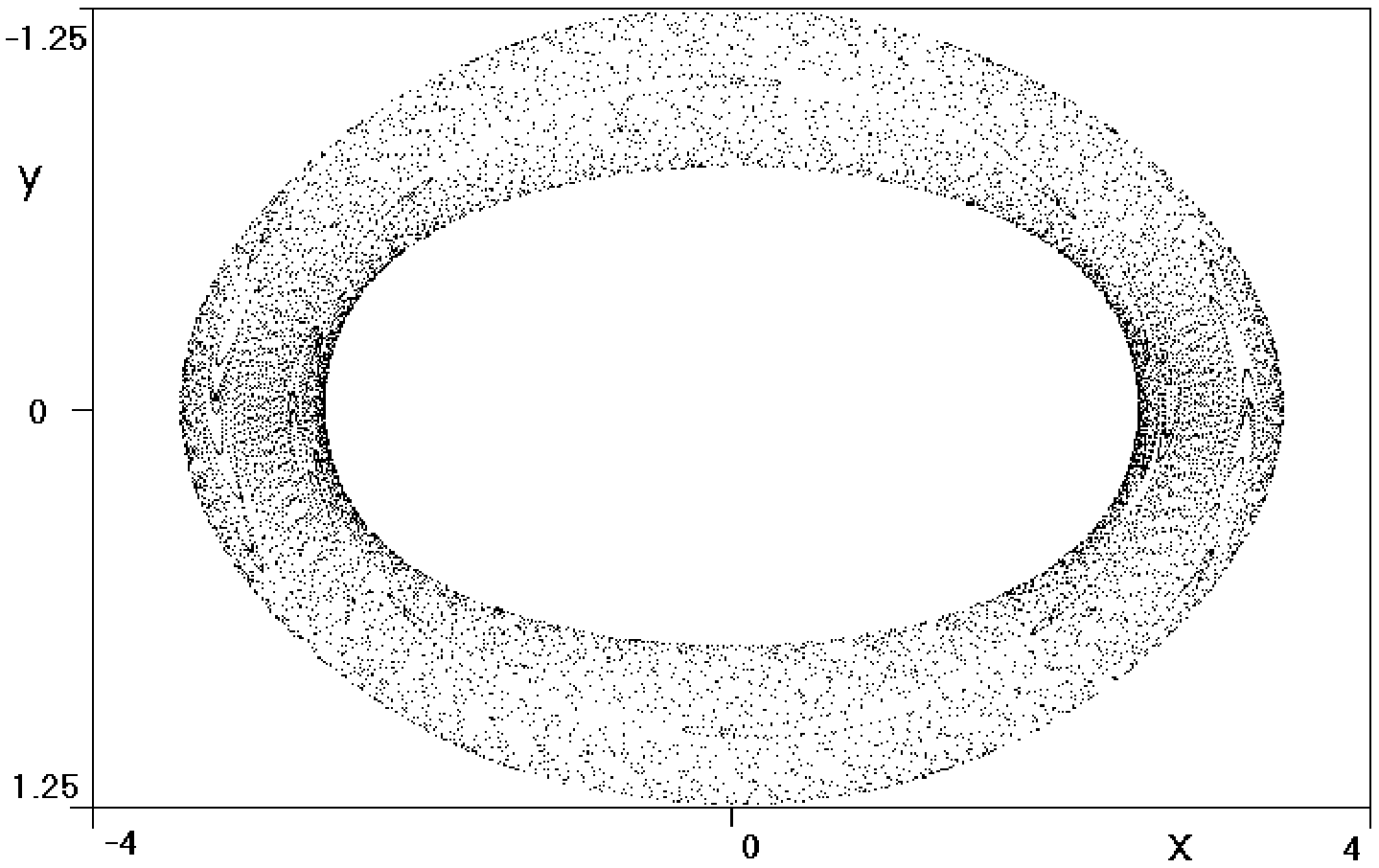}(b)}
 \caption{
 Irregular dynamics of the system with $a/c \ne
$const. Coordinates $(x,y)$ of a particle at the moments of collisions
with a wall are shown. Parameters of the system: $\eps=0.001,$
$\omega=0$, $A_1=0.15$, $A_2=0.25$, $d_1=3$, $d_2=1$, $\phi=\pi/3$. a)
Time of the integration is one-half of the slow period. b)Time of the
integration is equal to 2 slow periods.}
\end{figure*}

\end{document}